\documentclass[aps,prl,twocolumn,superscriptaddress]{revtex4-1}

\usepackage{graphicx}
\usepackage{amsmath} 
\usepackage{amssymb}	
\usepackage{hyperref}
\hypersetup{
    colorlinks=true,       
    linkcolor=black,          
    citecolor=black,        
    filecolor=magenta,      
    urlcolor=blue           
}

\begin{document}


\title{Strong Field Molecular Ionization in the Impulsive Limit: Freezing Vibrations with Short Pulses}


\author{P\'{e}ter S\'{a}ndor}
\author{Vincent Tagliamonti}
\author{Arthur Zhao}
\affiliation{Department of Physics and Astronomy, Stony Brook University, Stony Brook NY 11794-3800}

\author{Tam\'as Rozgonyi}
\affiliation{Institute of Materials and Environmental Chemistry, Research Centre for Natural Sciences, Hungarian Academy of Sciences, Budapest 1117 Magyar tud\'{o}sok krt. 2, Hungary}
\author{Matthias Ruckenbauer}
\author{Philipp Marquetand}
\affiliation{University of Vienna, Faculty of Chemistry, Institute of Theoretical Chemistry, W\"ahringer Str. 17, 1090 Wien, Austria}

\author{Thomas Weinacht}
\affiliation{Department of Physics and Astronomy, Stony Brook University, Stony Brook NY 11794-3800}

\date{12 February 2016}

\begin{abstract}
We study strong-field molecular ionization as a function of pulse duration. Experimental measurements of the photoelectron yield for a number of molecules reveal competition between different ionization continua (cationic states) which depends strongly on pulse duration.  Surprisingly, in the limit of short pulse duration, we find that a single ionic continuum dominates the yield, whereas multiple continua are produced for longer pulses.  Using calculations which take vibrational dynamics into account, we interpret our results in terms of nuclear motion and non-adiabatic dynamics during the ionization process.\\
\textit{Preprint of mansucript published in \href
       {http://dx.doi.org/10.1103/PhysRevLett.116.063002}{Phys. Rev. Lett., \textbf{116}, 063002 (2016)}.}
\end{abstract}


\maketitle

Strong-field molecular ionization plays an important role in the generation of attosecond pulses and electron wave packets \cite{Corkum_attosecond,Baumert_NJP,agostini2004physics,goulielmakis2010real}.  It can also be used to track excited state molecular dynamics and for molecular imaging \cite{Baumert1992488,Janssen_JChemPhys,PhysRevLett.106.073001,Villeneuve200569,Meckel13062008,murnane_br2_pnas}.  A detailed understanding of the ionization dynamics is crucial for developing these frontier areas of molecular science \cite{IS_StreakCam}. In particular, with a push to improve time resolution in molecular dynamics experiments \cite{Krausz_Nature1,PhysRevLett.102.213001}, generate multi-hole electron wave packets \cite{Gibson_JPhysB,PhysRevLett.95.013001,Kling14042006,McFarland21112008,bergues2012attosecond} and single attosecond pulses, it is important to understand how ionization depends on the duration of the strong-field driving pulse \cite{Sansone20102006, Chen10062014}.  Here, we study strong-field molecular ionization as a function of pulse duration, going from several tens of fs to below 10 fs, where vibrational dynamics is frozen out (the `impulsive limit') \cite{Xie2015}. Surprisingly, we find that as we shorten the pulse duration from about 40 fs to less than 10 fs, there is a dramatic change in the photoelectron spectrum, which reflects a change in the combination of ionic continua that are accessed during the ionization process. We observe similar behavior in three different molecules (CH$_2$IBr, CH$_2$BrCl and C$_6$H$_5$I) and demonstrate that the result depends more sensitively on pulse duration than spectral content.  For CH$_2$IBr, we further interpret the experimental measurements in terms of calculations of strong-field molecular ionization which include vibrational dynamics on intermediate neutral states during the ionization process.

Our experimental apparatus consists of an amplified Ti:sapphire laser system, which produces 30 fs pulses with an energy of 1 mJ and a central wavelength of 780 nm. The pulses are focused into an Argon gas cell to create a filament and broaden the spectrum \cite{stibenz2006self}. The pulses are compressed to near the transform limit with a 4-f grating compressor, and measured using a Self-Diffraction (SD) FROG apparatus \cite{trebino1997measuring}. The broadest spectrum we produce is capable of supporting sub 6 fs pulses, and FROG measurements place an upper limit on the duration of the full bandwidth pulses of about 8-9 fs. The spectrum is cut using a variable slit in the grating compressor in order to obtain the variable bandwidth for the measurements below. The spectrum of the pulse is adjusted at the focusing element instead of the Fourier plane in order to avoid hard cutoffs at the edges of the spectrum, which would lead to a structured pulse in the time domain.

The linearly polarized laser beam crosses an effusive molecular beam in a vacuum chamber. Here we generate electrons and ions, which are detected by accelerating them toward a dual stack of microchannel plates and phosphor screen with an electrostatic lens configured for velocity map imaging (VMI).  The VMI lens produces a two-dimensional projection of the three-dimensional charged particle velocity distribution \cite{parker:1997}. The hit locations on the phosphor screen from each laser shot are recorded and digitized by a CMOS camera which reads them into memory as separate images. The laser intensity is adjusted between 10-13 TW/cm$^2$ to keep the ionization yield roughly constant as the pulse duration is varied, yielding about 20$\pm$10 electrons per laser shot. A computer algorithm extracts the hit locations from each image and synthesizes a single data image for any given laser pulse parametrization. This image is inverse-Abel transformed with the BASEX method \cite{dribinski2002reconstruction} and converted to a photoelectron spectrum. We focus on the yield that is generated $\pm$30$^{\circ}$ around the laser polarization direction. Integrating over all angles yields similar results, with slightly less contrast of the peaks.

\begin{figure}[h!]
	\centering
	\includegraphics[width=0.5\textwidth]{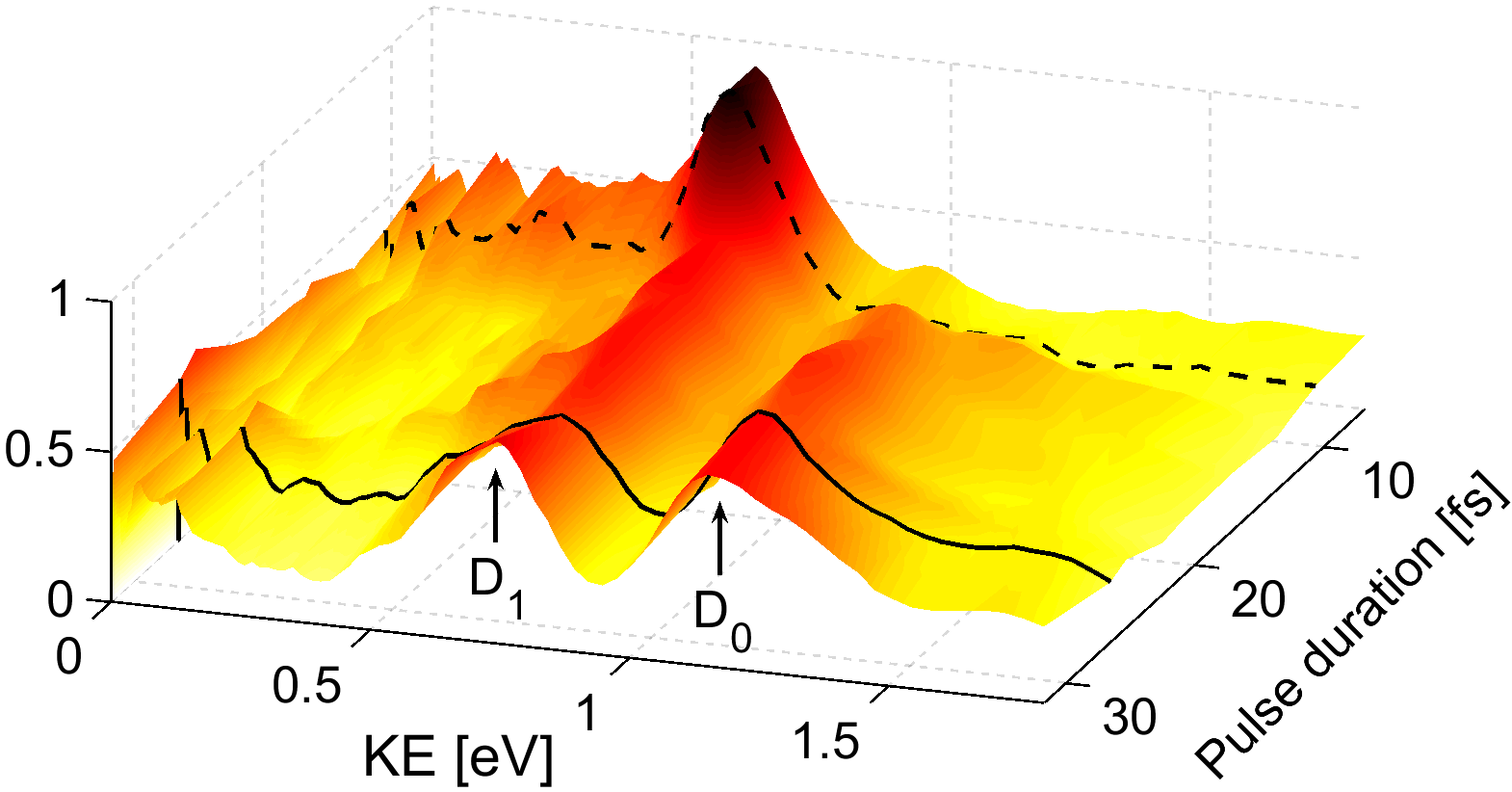}
	\caption{(color online) Photoelectron spectra (yield vs photoelectron kinetic energy, KE) for ionization of CH$_2$IBr for different pulse durations.}
	\label{fig:exp_only}
\end{figure}

Figure \ref{fig:exp_only} shows photoelectron spectra for CH$_2$IBr as a function of pulse duration. For longer pulses ($>$20 fs), two peaks are visible: one at $\approx$1.2 eV and the other at $\approx$0.7 eV. Earlier work assigned these peaks to leaving the molecule in the first two states of the molecular cation: D$_0$ and D$_1$, respectively. The assignment of the spectrum to the molecular ions was verified using electron-ion coincidence spectroscopy \cite{coincidence1}. While the yield from 0-0.2 eV and around 1.6 eV can be assigned to D$_2$/D$_3$ \cite{coincidence1}, we focus on the yield to D$_0$ and D$_1$ for simplicity here.  Earlier work \cite{SFImodeltheory} also established that these peaks involve resonance enhancement via intermediate neutral states that Stark shift into resonance during the ionization process \footnote{We note that intermediate neutral states can play an important role even for very short pulses, where the resonance condition is only met for a relatively short time, provided that there is sufficiently strong coupling.}. For longer pulses, ionization proceeds such that D$_0$ and D$_1$ are populated with roughly equal probability. However, as the pulse is shortened to below 12 fs, the yield for the D$_0$ peak diminishes and eventually becomes negligible compared to that of the D$_1$ peak.  This is surprising given that the ionization potential for D$_0$ (9.7 eV) is lower than for D$_1$ (10.2 eV) \cite{coincidence1}, and that the bandwidth of a shorter pulse is broader.

\begin{figure}[h!]
	\centering
	\includegraphics[width=0.4\textwidth]{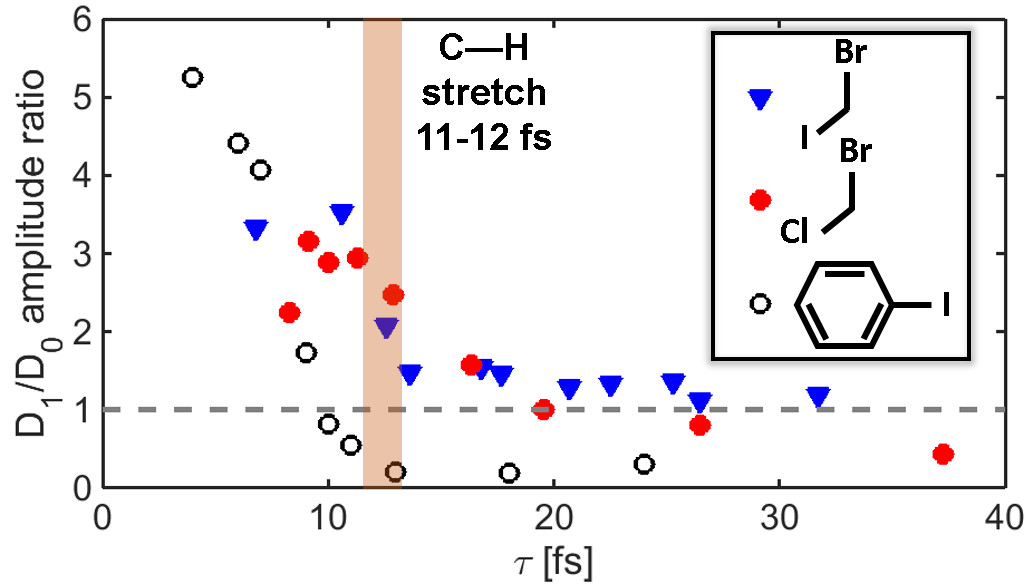}
	\caption{(color online) Ratio of D$_1$/D$_0$ as a function of pulse duration $\tau$ for three different molecules.}
	\label{fig:all_molecules}
\end{figure}

We carried out similar measurements for other molecules and observed similar dynamics.  Figure \ref{fig:all_molecules} shows the D$_1$ and D$_0$ ratio as a function of pulse duration for three different molecules: CH$_2$IBr, CH$_2$BrCl and C$_6$H$_5$I.  As the figure illustrates, all three molecules show similar behavior as a function of pulse duration. A shaded vertical bar marks the impulsive limit, corresponding to the C-H stretch vibrational period ($\approx$11 fs - the shortest vibrational period for organic molecules) \cite{JPhysChem:CH_stretch}.

We now aim to interpret the measurements and determine whether the mechanism underlying the switching/control is driven by dynamics or spectral content.  Since a shorter pulse duration requires a broader spectrum, it is natural to ask whether the suppression of ionization to D$_0$ is driven by new frequency components in the pulse, or rather by the pulse becoming shorter.

\begin{figure}[h!]
	\centering
	\includegraphics[width=0.4\textwidth]{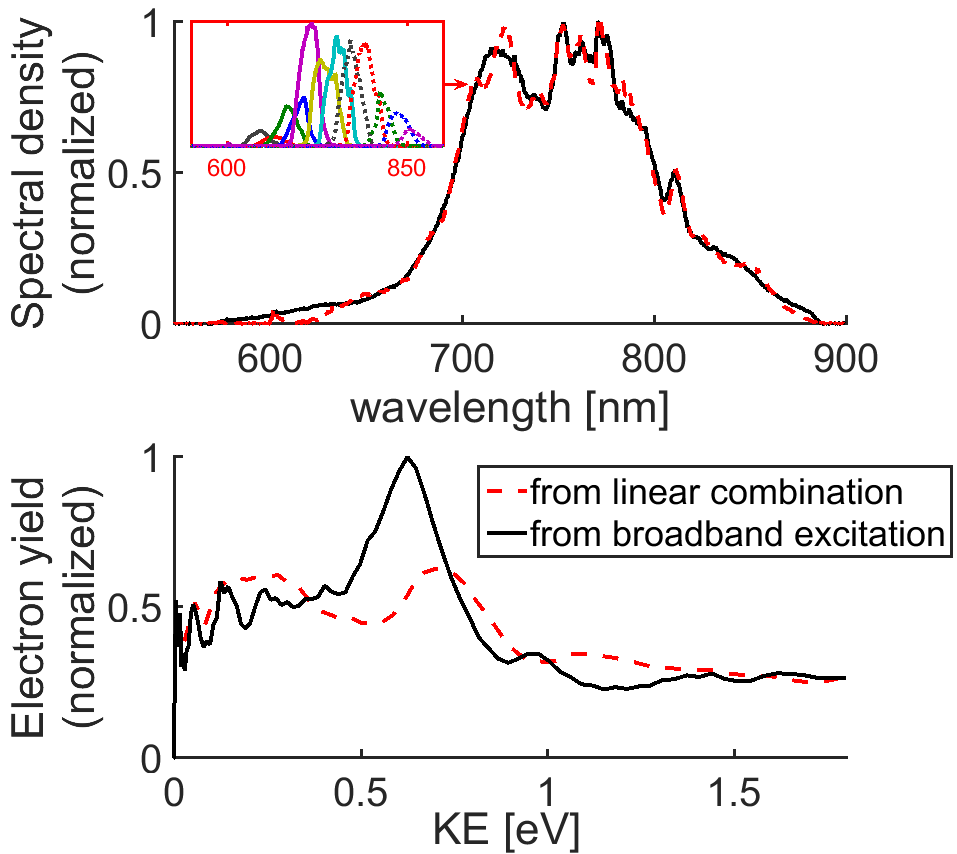}
	\caption{(color online) Top panel: Optical spectra for broadband pulse (solid black line) and the result of summing narrowband spectra (dashed red line). Bottom panel: photoelectron spectra of CH$_2$IBr acquired with full bandwidth optical spectrum (solid black line), and the result of forming a linear combination of photoelectron spectra each acquired with narrowband optical excitation (dashed red line). The latter are added in proportion to their spectral weights as shown in the inset and described in the text.}
	\label{fig:added_up}
\end{figure}

We first address this question by making measurements with a series of narrowband optical pulses with different central frequencies, adding up the photoelectron spectra with the appropriate weights and comparing the result with the photoelectron spectrum measured for a short pulse that includes all the spectral components coherently. This idea is illustrated in figure \ref{fig:added_up}. The  top panel shows the optical spectrum of the short pulse and the weighted sum of the narrow optical spectra together, while the lower panel shows the resulting photoelectron spectra - one curve for the sum of the photoelectron spectra produced with narrowband pulses, and one curve for the photoelectron spectrum produced by a broadband pulse. The photoelectron spectra for the narrowband pulses were added in proportion to the coefficients for the narrowband optical spectra in forming the broadband spectrum as a linear combination. While the optical spectra are almost identical, there are significant differences between the two photoelectron spectra, indicating that it is not a single frequency in the pulse spectrum which drives the switching between ionic continua.\\

A second test that we performed was to vary the pulse duration while keeping the spectral content the same.  This can be accomplished by placing a second-order spectral phase (chirp) on the broadband pulse, while varying the pulse energy to maintain a roughly constant yield. These results (to be presented in a forthcoming publication \cite{vincent_paper}) showed that the suppression of D$_0$ only takes place for a short pulse, corroborating the conclusion drawn above.

The observations described above suggest that there is some molecular dynamics which leads to both ionic states being populated, and if the pulse is shorter than the timescale for this dynamics, then only a single ionic state is populated.  As the photoelectron spectrum is determined at the moment the electron is born in the continuum (i.e. it is not affected by possible subsequent dynamics in the molecular cation), we argue that the dynamics leading to the selectivity must be neutral dynamics, involving an excited neutral state en route to the ionization continuum. As in earlier work which established the importance of dynamically Stark-shifted resonances \cite{Gibson_MPI,PhysRevLett.59.1092,PhysRevLett.69.1904} in strong-field molecular ionization \cite{PhysRevLett.69.1904,coincidence1,Freeman_control}, our current intensity and wavelength dependent measurements indicate that neutral Rydberg states Stark shift into resonance during the ionization process. The correlation between a neutral Rydberg state and low lying states of the molecular cation is typically large for only a single cationic state with a similar configuration of the core - i.e. the Dyson norm for a given neutral state is large for a single low lying state of the cation, and close to zero for other states \cite{PhysRevA.86.053406}. While Dyson correlations can be poor for low lying neutral states in strong-field ionization, they are better for higher lying states of the neutral where the electron which is removed during ionization does not interact with the ionic core very much and does not modify the core configuration.  This means that once an intermediate neutral Rydberg state comes into resonance, it typically couples to a single ionic continuum \cite{Freeman_control}.  Thus, for resonance-enhanced ionization to multiple continua, as is the case for a $\approx$40 fs pulse, multiple intermediate states must be involved in the ionization dynamics.

Our earlier work considered resonance enhanced ionization with separate uncoupled intermediate states for each ionization continua \cite{Freeman_control}. We extend this model to include coupling between the intermediate states, as our new measurements suggest that separate uncoupled intermediate resonances cannot account for the pulse duration dependence we observe. If the bandwidth associated with different pulse durations were to select between different independent resonances, then one would expect to find a single ionic continuum favored for a long pulse (narrow bandwidth) rather than for a short pulse (broad bandwidth), since a shorter pulse contains a larger bandwidth, which would provide less selectivity between separate resonances.   Furthermore, frequency-dependent measurements of the ionization yield such as the ones illustrated in figure \ref{fig:added_up} indicate that when there is resonance enhancement of the ionization yield, then it is through a single neutral state correlated with D$_1$.  These measurements are discussed in detail in a separate publication \cite{vincent_paper}.

While in principle both laser-driven resonance \cite{Janssen_JChemPhys} and non-adiabatic dynamics could be responsible for coupling excited states, given the frequency-dependent measurements shown in figure \ref{fig:added_up}, and motivated by earlier work \cite{DomckePEScalc}, we focus on non-adiabatic dynamics as an explanation for the measurements shown in figures \ref{fig:exp_only} and \ref{fig:all_molecules}.  We carry out calculations for CH$_2$IBr that include non-adiabatic coupling between excited states which support the idea that molecular dynamics drives the switching between continua as a function of pulse duration.

Before modeling the strong-field ionization with numerical integration of the time-dependent Schr\"{o}dinger equation (TDSE), we carry out \textit{ab initio} electronic structure calculations at the MS-CASPT2 level of theory \cite{MSCASPT2} in order to determine which electronic states play a crucial role in the ionization process. Details on the electronic structure calculations are given in the supplementary information. The strong field ionization simulations are based on a simple model \cite{SFImodeltheory} which includes Stark shifted intermediate neutral resonances. This model is now extended to include vibrational dynamics and non-adiabatic coupling between multiple intermediate neutral states, as considered in earlier calculations for weak(perturbative) laser fields \cite{DomckePEScalc}. We focus on CH$_2$IBr, for which we made the most detailed measurements and calculations.

\begin{figure}[h!]
	\centering
	\includegraphics[width=0.5\textwidth]{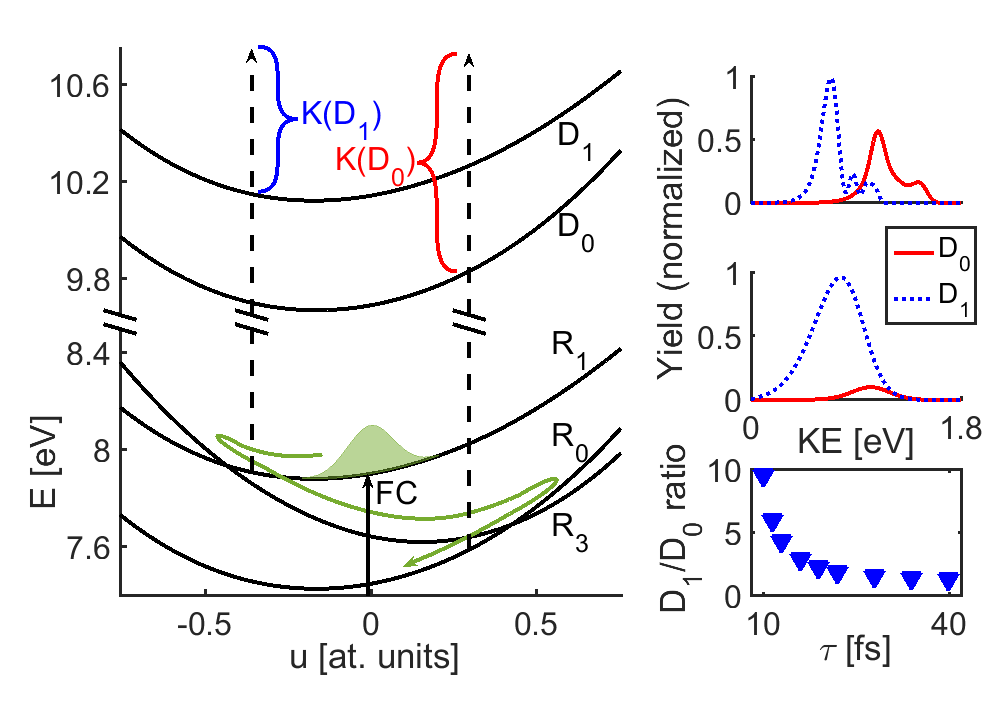}
	\caption{(color online) Left panel: calculated potential energy surfaces for CH$_2$IBr along the CH$_2$ wagging mode. FC: Franck-Condon point for excitation from the minimum of S$_0$ (u=0). Upper right panel: calculated photoelectron spectra for 40 fs pulse. Middle right panel: calculated photoelectron spectra for 10 fs pulse. Bottom right panel: calculated D$_1$/D$_0$ yield ratio as a function of pulse duration $\tau$.}
	\label{fig:theory_only}
\end{figure}

As prior measurements suggest that resonance enhancement occurs at the five-photon level \cite{SFImodeltheory}, we considered Rydberg states (R$_0$, R$_1$ and R$_3$) $\approx$8 eV above the ground state which are correlated (i.e. similar electronic configurations) with the low-lying ionic states (D$_0$, D$_1$ and D$_3$), and whose coordinate dependence follows those of the ionic states with which they are correlated.  We then considered whether any nuclear coordinates led to coupling between these states. While the potential energy curves of the Rydberg states around 8 eV are largely parallel as a function of most vibratonal coordinates, we found one mode (CH$_2$ wagging) along which motion leads to degeneracy (and therefore to population transfer via non-adiabatic coupling) between states correlated with D$_0$, D$_1$ and D$_3$.  The potential energy curves of these states along this normal mode coordinate are shown in Figure \ref{fig:theory_only}. In our calculations, population excited to R$_1$ (which based on experimental measurements is most strongly coupled to S$_0$ via the laser \cite{vincent_paper}) can quickly relax to R$_3$ and R$_0$ via rapid nuclear dynamics and non-adiabatic coupling.  Based upon matches of the computed energy differences and similarities between electronic configurations of the electronic configurations, R$_0$, R$_1$ and R$_3$ are coupled to D$_0$, D$_1$ and D$_3$ respectively.
The strong-field ionization calculations produced the photoelectron spectrum as a function of pulse duration, as in the measurements.  The energies of the resonant intermediate states were based on the electronic structure calculations and comparison with experimental spectra.  Laser parameters, such as the intensity, central frequency and pulse duration, were based on experimental parameters.  The coupling strengths (multiphoton Rabi frequencies) are given in the supplementary material. 

As the S$_0\rightarrow$R$_1$ resonance dominates, population is initially transferred from S$_0$ to R$_1$. As figure \ref{fig:theory_only} illustrates, motion along the CH$_2$ wagging mode couples states R$_0$, R$_1$ and R$_3$. The Franck Condon point (minimum of S$_0$) is close to the R$_1$/R$_3$ crossing, leading to rapid population transfer from R$_1$ to R$_3$ ($\lessapprox$5 fs). Within $\approx$10 fs the wave packet on R$_3$ can proceed to the R$_3$/R$_0$ crossing.  Thus, for a long pulse, ionization can proceed to a mixture of the ionic states D$_0$, D$_1$ and D$_3$ coupled with the three neutral states R$_0$, R$_1$ and R$_3$.   While our measurements show evidence of ionization to all three of these states, we concentrate on the competition between D$_0$ and D$_1$ because the measurements are cleanest for these states.  In the limit of a short laser pulse, one might expect D$_1$ (which is correlated with R$_1$) to dominate the ionization yield, since R$_1$ can shift into resonance and there is not sufficient time for the wave packet to move away from the FC on R$_1$ during the ionization.  The ionization calculations aimed to test this hypothesis.

As the top right and middle panels of figure \ref{fig:theory_only} illustrate, the photoelectron spectrum for a long pulse (40 fs) shows peaks corresponding to D$_0$ and D$_1$, whereas the photoelectron spectrum for a 10 fs pulse shows a single peak corresponding to D$_1$ only.  This is in agreement with the results shown in figure \ref{fig:exp_only}, which shows two peaks corresponding to D$_1$ and D$_0$ for a long pulse and a single peak corresponding to D$_1$ for a short pulse. The bottom right panel shows a decreasing D$_1$/D$_0$ ratio as a function of pulse duration, in agreement with the results shown in figure \ref{fig:all_molecules}.  One aspect of the measurements which is not reflected in the calculations is the width of the peaks in the PES as a function of pulse duration.  The measurements show relatively narrow peaks for both short and long pulse durations, while the calculations show peaks which broaden as a function of decreasing pulse duration.

Our interpretation of the pulse duration dependence relies on neutral state resonances enhancing the ionization yield.  Thus one would expect that there is no change in the ionization yield for different ionic continua with pulse duration if there are no important resonances.  In order to test this, we performed measurements of the photoelectron spectrum vs pulse duration in CS$_2$, for intensities where there are no intermediate resonances for our laser frequency.  We also performed measurements in CH$_2$IBr for very low intensities where the intermediate states do not Stark shift into resonance.  In both cases we found that the photoelectron spectrum did not change substantially with pulse duration, as one would expect based on our interpretation which relies on dynamics in intermediate neutral states.

In conclusion, we study the state-resolved ionization yield as a function of pulse duration for several molecules and find that for relatively long pulses, vibrational dynamics and non-adiabatic coupling between resonant intermediate states play an important role. For impulsive ionization with pulses less than 10 fs in duration, vibrational dynamics is frozen and no longer plays an important role in the ionization process.  The transition between the two regimes is clearly visible in the photoelectron spectrum, in which a dramatic switching between single and multiple continua is observed.  Surprisingly, resonance-enhanced ionization plays an important role for even the shortest pulses, containing nearly an octave optical bandwidth.  Our results suggest that these considerations should be relevant for a broad range of molecules in strong laser fields.
\begin{acknowledgments}
This work has been supported by the National Science Foundation under award number 1205397 and the Austrian Science Fund (FWF) through project P25827. Support from the European XLIC COST Action 1204 is also acknowledged.
\end{acknowledgments}


\bibliographystyle{apsrev_nourl}
\vspace*{3.5cm}
\noindent{\Large{\textbf{Supporting Information}}}

\section{Experimental Pulse Characterization}
\begin{figure}[h!]
\centering
\includegraphics[width=0.5\textwidth]{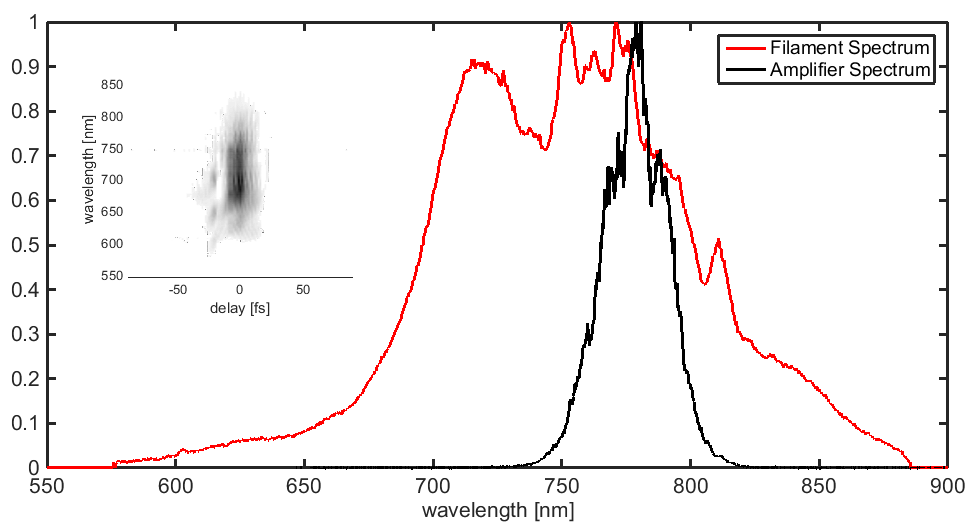}
\caption{Optical spectra for the amplified beam (input to Argon gas cell) and the filament (output of gas cell). Inset: Self-Diffraction FROG trace of a dispersion-compensated filamentation pulse.}
	\label{spectrum}
\end{figure}

An important aspect of our work is establishing that the pulses used in the experiments are Fourier limited, such that the pulse duration inferred from a measurement of the optical spectrum reflects the actual pulse duration.  We characterized the pulses using self diffraction frequency resolved optical gating (FROG) \cite{trebino1997measuring}.  This allowed us to establish an upper limit on the pulse duration as noted in the manuscript.  The FROG measurement along with the optical spectrum is shown in figure \ref{spectrum}.
\section{Ab initio calculations}

The vibrational normal mode coordinates of the neutral CH$_2$BrI in its ground electronic state were determined with density funtional theory using the Gaussian program package~\cite{Gaussian} with the B3LYP functional~\cite{B3LYP} and aug-cc-pVTZ basis for the C, H and Br atoms and aug-cc-pVTZ-PP basis for the I atom~\cite{augccpVTZ}. The same level of theory had been applied to determine the equilibrium geometry before (see Ref.~\cite{PCCP}). One-dimensional potential energy curves were then computed along these coordinates for the lowest five electronic states of the ionic CH$_2$BrI$^+$ at the multistate CASPT2 (MS-CASPT2) level of theory~\cite{CASPT2_1,CASPT2_2} as implemented in the Molcas~8.0 program package~\cite{Molcas} using the Douglas-Kroll-Hess Hamiltonian~\cite{DKH} and the ANO-RCC basis~\cite{ANO_1,ANO_2} set with the following contractions: 3s2p1d for H, 4s3p2d1f for C, 6s5p3d2f1g for Br and 7s6p4d3f2g for I atoms. These are equivalent to the triple-zeta+polarization type basis. The basis set was augmented by the Rydberg.Kaufmann.8s8p8d.8s8p8d diffuse orbitals~\cite{Rydberg} centered at the center of mass of the molecule in order to describe Rydberg electronic states. The active space consisted of 10 active orbitals with 12 or 11 electrons for the neutral molecule and the cation, respectively. The active orbitals were the two bonding and two antibonding carbon-halogen $\sigma$ orbitals, the four halogen lone pair orbitals and two Rydberg orbitals: one with a' and another with a" symmetry. (In the following we refer to these two Rydberg orbitals indicating also their symmetries as Rydb.a', Rydb.a", respectively.) Five doublet electronic states were included in the MS-CASPT2 computations for the cation, while 18 singlet and 17 triplet states were considered in the computations for the neutral species. The spin-orbit coupling (SOC) was computed for both the ionic and the neutral states using atomic mean field integrals (AMFI) \cite{AMFI}. According to their energies and to the characters of their excited Rydberg orbitals the obtained 32 Rydberg spin-orbit eigenstates can be divided into two main groups: one main group is characterized by excitations from halogen lone pair orbitals to the Rydb.a' orbital and the other main group is characterized by excitations from halogen lone pair orbitals to the Rydb.a" orbital. As there are 4 different halogen lone pair orbitals, both main groups contain 4 subgroups of 4 almost degenerate spin-orbit eigenstates. The deviation of state energies within each subgroup is less than 0.04 eV. Therefore, in our computations each selected subgroup of Rydberg states and each degenerate pair of ionic states were represented by a single state, respectively.
Table~I shows the energy gaps between the subgroups of Rydberg states separately for the two main groups: The second column refers to the main group of Rydberg states characterized by excitations to the Rydb.a' orbital while the third column refers to the main group of Rydberg states characterized by excitations to the Rydb.a" orbital. The fourth column shows excitation energies between the pair of degenerate ionic states.
\begin{table}[h!]
\caption{\label{cv} Energy gaps, $\Delta$E, in eV between subgroups of Rydberg states characterized by the same excited Rydberg orbitals and energy gaps between ionic states, D$_j$ at the Franck-Condon point ($u=0$).}
\begin{tabular}{lcccccc}
 \hline
 \hline
  j &  $\Delta$E(n$_j \rightarrow $Rydb.a') && $\Delta$E(n$_j \rightarrow $Rydb.a") &&  $\Delta$E(D$_0 \rightarrow $D$_j$)  \\ \hline
0  &  0   &&  0   && 0 \\
1  & 0.38 && 0.44 && 0.44  \\
2  & 0.98 && 0.98 && 1.00  \\
3  & 1.13 && 1.14 && 1.16  \\ \hline
 \hline
\end{tabular}
\end{table}
The numerical results in Table~I show that the energy gaps within a group of Rydberg states almost perfectly match the energy gaps found between the ionic states. In addition, the shapes of the potential energy curves of the Rydberg states along the selected vibrational mode also match very well with the shapes of the corresponding ionic potential curves. These suggest that a unique correlation exists between the groups of Rydberg states and ionic states and thus the different groups of Rydberg states are expected to enhance ionization selectively to one degenerate pair of ionic states. We therefore selected those Rydberg states for our dynamics, which are the closest to resonance with 5 photon excitation from neutral ground state and are supposed to play the most important role in the dynamics. These are two subgroups of n$\rightarrow$Rydb.a" type Rydberg states: one subgroup correlated to D$_0$ and the other correlated to D$_1$ states. We model these two subgroups of Rydberg states by model states R$_0$ and R$_1$, respectively. The third neutral excited state supposed to play a crucial role in the dynamics is an n$\rightarrow$Rydb.a' type subgroup of Rydberg states which is neither correlated to D$_0$ nor D$_1$, but to D$_3$ and which crosses both the R$_0$ and R$_1$ states along the CH$_2$ wagging mode close to the Franck-Condon point, as indicated in Fig.4 of the paper. We model this subgroup of Rydberg states by the model state R$_3$ and perform one-dimensional dynamics along the CH$_2$ wagging mode "connecting" the R$_0$ and R$_1$ potentials by the R$_3$ potential. There are Rydberg states of n$\rightarrow$Rydb.a' type correlated to the D$_2$ ionic states also close to the 5 photon resonance level, but these Rydberg states together with the corresponding D$_2$ ionic state are neglected in the model, since they are considered to play a negligible role in the dynamics and in the ionization process.

\section{Wavepacket dynamics}

Ionic potentials for D$_0$, D$_1$ and D$_3$ along the CH$_2$ wagging mode were fitted separately by harmonic functions. As these potentials fit well to the corresponding R$_0$, R$_1$ and R$_3$ states, respectively, when their offset is adjusted, we used these harmonic potentials with adjusted offsets as diabatic potentials for the neutral excited states as well. The reduced mass along the CH$_2$ wagging mode - as obtained from the DFT computation - is 1.102~amu and the harmonic vibrational frequency of this mode in the neutral ground state is 1176~cm$^{-1}$. The harmonic frequencies for the D$_0$, D$_1$ and D$_3$ states as obtained from the fitting were 1200, 1130 and 1260 cm$^{-1}$, respectively. The positions of the potential minima were -0.15~au for D$_0$, D$_1$, R$_0$ and R$_1$ and +0.15~au for D$_3$ and R$_3$. We adjusted the offset of the ground ionic state at the Franck-Condon point ($V_0^{ion}(u=0)$) to the experimental ionization potential of 9.69~eV~\cite{Baer}.
Our ab initio computation of the lowest Rydberg state reported in \cite{Penner1990} overestimated the experimental energy by 0.3 eV. Since on the other hand the energy gaps between the Rydberg states match the gaps between the ionic states, we corrected the offset for all the applied Rydberg state potentials by the same 0.3 eV value (i.e., we shifted the potentials downwards by 0.3 eV.) Furthermore, we introduced non-adiabatic couplings between the R$_1$/R$_3$ and between the R$_3$/R$_0$ neutral states. The coupling functions, $V_{e,e'}(u)$ ($e$,$e'$  = R$_0$, R$_3$ and $e$,$e'$ = R$_1$, R$_3$) had Gaussian shapes with equal amplitudes of 0.2 THz and full widths at half maximum of 0.6 au and were centered at the intersection points.

We assumed that the highest kinetic energy ($E_\text{max}$) of a photoelectron is 2 eV and that the neutral-to-ionic multiphoton couplings, $\chi_{j \alpha}(t)$ are independent of the kinetic energy, $E$, of the emitted electron for $E < E_\text{max}$ and are zero otherwise. These approximations lead to  simplified equations of motion when wavefunctions $\Psi_\alpha^\text{ion} (E,u,t)$ of ionic state D$_\alpha$ are expanded in terms of Legendre polynomials $P_l(E)$ of $E$ as~\cite{DomckePEScalc,DIC_2,SFImodeltheory}
\begin{equation}
\Psi_\alpha^\text{ion}(u,E,t) = \sum_{l=0}^{M\rightarrow\infty}\Phi^{(\alpha)}_{l+1}(u,t) \sqrt{\frac{2l+1}{E_\text{max}}}P_l\left (\frac{2E}{E_\text{max}} -1 \right ) \label{eqn:expansion}
\end{equation}
\begin{equation*}
\textrm{with } \alpha = \text{D}_0, \text{D}_1, \text{D}_3 \label{eqn:expansion2}
\end{equation*}
Eliminating the off-resonant electronic states adiabatically in the same way as done in Ref.~\cite{ADEL,SFImodeltheory}, and inserting the potential-type non-adiabatic couplings, $V_{e,e'}$ between resonant neutral excited states into the time-dependent Schr\"{o}dinger equation, the equations of motion take the following form:
\begin{align}
i\hbar\dot{\Psi}_g =& \left ( \hat{T} + V_g(u) \right ) \Psi_g + \sum_{e} \chi_{eg}^*(t)\Psi_e \nonumber \\
i\hbar\dot{\Psi}_e =& \left ( \hat{T} + V_e(u) +  \omega^{(s)}_e(t) \right ) \Psi_e + \chi_{eg}(t)\Psi_g   \nonumber \\
&+ \sum_{e' \neq e} V_{e,e'} \Psi_{e'} + \sum_{\alpha} \chi_{\alpha e}^*(t)\Phi_1^{(\alpha)} \nonumber\\
i\hbar\dot{\Phi}_1^{(\alpha)} =& \left ( \hat{T} + V_\alpha^{ion}(u) + \frac{E_\text{max}}{2} + U_p(t) \right )\Phi_1^{(\alpha)} \nonumber\\
&+ \frac{1}{\hbar}\rho_2 \Phi_2^{(\alpha)}  +  \sum_{e} \chi_{\alpha e}(t)\Psi_e\nonumber\\
i\hbar\dot{\Phi}_k^{(\alpha)} =& \left ( \hat{T} + V_\alpha^{ion}(u) + \frac{E_\text{max}}{2} + U_p(t) \right ) \Phi_k^{(\alpha)} \nonumber\\
&+ \rho_k \Phi_{k-1}^{(\alpha)} + \rho_{k+1} \Phi_{k+1}^{(\alpha)} ~, \label{eqn:final}
\end{align}
The above equations can be regarded as straightforward generalization of the equations~8 of Ref.~\cite{SFImodeltheory} under the assumptions that no excited states other than those considered explicitly here come into resonance during the wavepacket motion.
In the above equations $\Psi_g(u,t)$ is the vibrational wavefunction in the ground electronic state, S$_0$ of the neutral molecule, $\Psi_e(u,t)$ is the wavefunction of the resonant neutral excited states, R$_e$. The $\rho_{k}$'s are constants defined as $\rho_{k}=\frac{(k-1)E_\text{max}}{2\sqrt{4(k-1)^2-1}}$.
The $U_p(t)$ is the ponderomotive potential being proportional to the intensity of the laser pulse. $\hat{T}$ is the kinetic energy operator, $V_e(u)$ is the diabatic potential energy curve of neutral electronic state R$_e$ and $V_\alpha^\text{ion}(u)$ is the potential energy curve (i.e., the field-free ionization potential) of ionic state D$_\alpha$. The $\omega_e^{(s)}(t)$ denote the dynamic Stark shifts of the neutral excited states, R$_e$. They are considered to be equal to the ponderomotive shift, $U_p(t)$ of the ionic states. The Stark shift of the neutral ground state and that of the ionic states were taken to be zero.

We assumed coordinate-independent Rabi-frequencies, i.e., that the multiphoton couplings, $\chi_{jm}(t)$ have the form of $\chi_{jm}(t)=\chi_{jm}^{(0)} [\varepsilon(t) e^{-i \omega_0 t}]^{N_{jm}}$, where $\chi_{jm}^{(0)}$ is a constant parameter and $\varepsilon (t)$ is the slowly varying envelope of the real electric field, $\epsilon(t)$ of the laser pulse with central frequency, $\omega_0$:
\begin{align}
\epsilon(t) = \frac{1}{2} \left ( \varepsilon(t) e^{-i \omega_0 t} +  \varepsilon^*(t) e^{+i \omega_0 t} \right ) \label{eqn:field}
\end{align}
The $N_{jm}$ is the photon order of the $j \rightarrow m$ transition, i.e., in our model $N_{g e} = 5$ and $N_{e \alpha} = 2$.  In order to focus on the effect of the non-adiabatic dynamics on the photoelectron spectrum, we have chosen the Rabi frequencies so that $\chi^{(0)}_{g1}$ be much larger than $\chi^{(0)}_{g0}$ and $\chi^{(0)}_{g3}$.  The amplitudes of the nonzero multiphoton couplings are given in Table~II for a reference intensity of 12~TW$/$cm$^2$.
\\

\begin{table}[h!]
\caption{\label{cv2} Amplitudes of the multiphoton couplings in THz for 12~TW$/$cm$^2$ intensity.}
\begin{tabular}{cccc | cccc}
 \hline
 \hline
   Transition  &&   $\chi_{ge}$   && Transition          && $\chi_{e\alpha}$ \\ \hline
  S$_0$ $\rightarrow$ R$_0$  &&  1 && R$_0$ $\rightarrow$ D$_0$ && 60    \\
  S$_0$ $\rightarrow$ R$_1$  && 20 && R$_1$ $\rightarrow$ D$_1$ && 20    \\
  S$_0$ $\rightarrow$ R$_3$  &&  1 && R$_3$ $\rightarrow$ D$_3$ && 20    \\ \hline
 \hline
\end{tabular}
\end{table}
Simulations were performed starting from the ground vibrational wavefunction of the neutral ground electronic state, S$_0$ using intensities corresponding to the experiments. Photoelectron spectra, $PES_{\alpha}(E)$ were computed separately for the ionic states $\alpha$ at the final time, $t_f$ of the simulation, when the pulse was over:
\begin{align}
PES_{\alpha}(E) = \int | \Psi_\alpha^\text{ion} (E,u,t_f) |^2 du  \label{eqn:PES}
\end{align}
To get smooth spectra and avoid artificial back-reflection of continuum population into the neutral states we applied $M=100$ continuum ladder coefficients, $\Phi_k$. In the simulations transform limited pulses were considered with central frequency, $\omega_0$ corresponding to the experimental photon energy of 390~THz. The temporal intensity profile of the laser pulse had a Gaussian shape with peak amplitude and full width at half maximum also  adjusted to experiment.

\end{document}